\documentclass[twocolumn,showpacs,prl,amsmath,amssymb]{revtex4}
\usepackage{graphicx}
\usepackage{dcolumn}
\usepackage{bm}

\begin{document}
\title{Anharmonic phonon frequency shift in MgB$_{\bm 2}$ }

\author{Michele Lazzeri, Matteo Calandra, and Francesco Mauri}
\affiliation{Laboratoire de Min\'eralogie-Cristallographie de Paris,
4 Place Jussieu, 75252, Paris cedex 05, France}

\date{\today}

\begin{abstract}

We compute the anharmonic shift of the phonon frequencies in MgB$_2$,
using density functional theory.
We explicitly take into account the scattering between different phonon modes
at different q-points in the Brillouin zone.
The shift of the E$_{2g}$ mode at the $\Gamma$ point is $+5 \%$ of the harmonic
frequency. This result comes from the cancellation between the
contributions of the four- and three-phonon scattering,
respectively $+10 \%$ and $-5 \%$.
A similar shift is predicted at the A point, in agreement with inelastic X-ray
scattering phonon-dispersion measurements.
A smaller shift is observed at the M point.

\end{abstract}
\pacs{63.20.Dj, 63.20.Kr, 71.15.Mb}

\maketitle

The discovery of high critical temperature (T$_c=39$K) in MgB$_2$ 
~\cite{Nagamatsu} has challenged our understanding of electron-phonon
coupling mediated superconductivity.
Several mechanisms might cooperate in determining the actual value of T$_c$
in MgB$_2$, including the existence of a double gap structure
~\cite{Shulga,Choi0,Golubov}, and anharmonic effects
~\cite{Liu,Kortus,Yildirim,Choi,Kunc,Boeri}.
Much attention has been devoted to the study of the E$_{2g}$ phonon mode, 
which consists of an antiphase vibration of the
two boron atoms, parallel to the hexagon plane.
This mode has the largest coupling with electrons according to inelastic X-ray
scattering measurements~\cite{Shukla},
in agreement with a commonly accepted interpretation of the MgB$_2$
electronic band structure~\cite{Kong,Liu,An}.
The E$_{2g}$ mode has also been associated with a supposedly large
anharmonicity, but a quantitative determinations of the anharmonic effects has
led so far to controversial results
~\cite{Shukla,Choi,Kortus,Liu,Yildirim,Kunc,Boeri}.

In particular, actual theoretical calculations indicate the anharmonic
frequency shift of the $E_{2g}$ mode at $\Gamma$ to be large, with varying
estimates that range from $+15\%$~\cite{Kortus,Liu},
up to $+20/25\%$~\cite{Yildirim,Choi} of the theoretical harmonic frequency
($\sim 65$ meV).
A comparably large shift is expected to affect this
mode all along the $\Gamma$-A direction~\cite{Kortus,Liu}.
Raman measurements~\cite{Bohnen,Hlinka,Quilty,Martinho}
seem to confirm the
prediction of a large anharmonicity at $\Gamma$.
In fact, the best resolved spectra~\cite{Quilty},
present a peak at $77$ meV ({\it i.e.} $18\%$ higher than the harmonic
frequency) that could be attributed to the E$_{2g}$ mode at $\Gamma$.
However, the structure observed below $40$ meV~\cite{Quilty}, and the clearly
asymmetric profile (reminiscent of a Fano resonance)
of the $77$ meV peak indicate that the Raman experiment
probes not just phonon vibrations, but electronic excitations as well.
Thus, the position of the $77$ meV peak
does not necessarily correspond to the E$_{2g}$ phonon frequency.
Moreover, Raman spectra display a drastic dependence on the temperature.
The width of the $77$ meV peak is $\sim 20$ meV at T$=40$K and reaches
almost $40$ meV at room temperature~\cite{Quilty,Martinho}.
This behavior has been tentatively attributed to
anharmonicity~\cite{Quilty,Martinho},
but, according to the calculations of Ref.~\cite{Shukla}, the E$_{2g}$-$\Gamma$
anharmonic broadening at room temperature is negligible, just $1.2$ meV.
The determination of the phonon frequency from Raman data requires
further analysis, as suggested in Ref.~\cite{Quilty}, together with a clear
understanding of the temperature dependence.

On the other hand, the prediction of a strong anharmonic frequency shift is
in stark contrast with the inelastic X-ray measurements of the MgB$_2$ phonon
dispersion of Ref.~\cite{Shukla}, which E$_{2g}$ phonon frequencies
measurements are in agreement (within $5\%$) with the
calculated harmonic phonon frequencies~\cite{Kong,Bohnen,Shukla} at several
points along the $\Gamma$-A direction.
Contrary to Raman scattering~\cite{Quilty,Martinho}, in X-ray inelastic
scattering the cross-section of electronic excitations is
much smaller than that of the phonon excitations.
Thus, X-ray measurements are a direct probe of the phonon frequencies.
Small anharmonic frequency shift is also suggested by a recent
point-contact spectroscopy measurement~\cite{Naidyuk}.

The origin of the discrepancy between the predicted large anharmonic shift of
the E$_{2g}$ mode~\cite{Kortus,Liu,Yildirim,Kunc,Boeri,Choi} and the inelastic
X-ray scattering measurements~\cite{Shukla}, might be traced back to the
approximations involved in the calculations.
We recall here that
the anharmonic frequency shift of a phonon mode $({\bm q}j)$ (identified by the
reciprocal-space vector ${\bm q}$ and by the band index $j$) is due to the
anharmonic interaction with the ensemble of all phonons, having different
momentum ${\bm q}$ and different $j$.
Up to now, theoretical studies of MgB$_2$ have been based on the study of a
frozen-phonon energy profile.
This approach completely neglects: (i) the dependence of the
anharmonic scattering on the exchanged phonon momentum ${\bm q}$, and (ii)
the coupling between phonons having different $j$.
The calculation of the anharmonic scattering between phonons having different
${\bm q}$ and different $j$ is possible, using first-principles~\cite{DFT}
techniques.
{\it E.g.}, it has allowed allowed the determination of
anharmonic line shift and broadening of the Raman modes in covalent
semiconductors~\cite{Debernardi}.
However, this approach is computationally demanding and has not been yet
applied to MgB$_2$.

In this Letter we compute the anharmonic frequency shift of the phonon
E$_{2g}$ mode of MgB$_2$ from first-principles~\cite{DFT}.
We explicitly take into account the scattering between different phonon modes
at different ${\bm q}$-points in the reciprocal space.
We use the density-functional perturbation theory of Ref.~\cite{DFPT}.
The anharmonic terms of the phonon Hamiltonian are obtained using a method,
based on the ``$2n+1$'' theorem, which allows the efficient calculation of
third order derivatives of total energy, at any point in the reciprocal space,
in metallic systems~\cite{Lazzeri02}.

\begin{figure}
\centerline{\includegraphics[width=85mm]{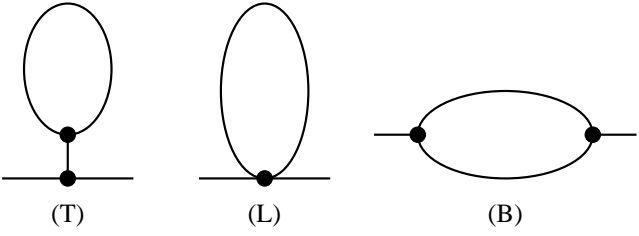}}
\caption{Diagrammatic representation of the leading terms in the perturbative
expansion of a non-harmonic phonon hamiltonian (see the text).}
\label{fig1}
\end{figure}

The phonon excitations of a crystal
are determined by the interatomic energy, and correspond to the
poles of the interacting phonon Green function~\cite{manybody,Menendez}.
$\Pi_{{\bm q}j}(\omega)$ represents the anharmonic contribution to the phonon
self-energy of the phonon mode $({\bm q}j)$ having frequency
$\omega_{{\bm q}j}$.
In the case $ |\Pi_{{\bm q}j}| \ll \omega_{{\bm q}j}$, the frequency shift
$\Delta_{{\bm q}j}$ and line broadening $\Gamma_{{\bm q}j}$, can be
obtained as
$\Delta_{{\bm q}j} = \mathcal{R}_e[\Pi_{{\bm q}j}(\omega_{{\bm q}j})]$
and
$\Gamma_{{\bm q}j} = -\mathcal{I}_m[\Pi_{{\bm q}j}(\omega_{{\bm q}j})]$,
$\mathcal{R}_e$ and $\mathcal{I}_m$ being the real and imaginary part
of a complex number.
The three lowest order terms in the perturbative expansion,
{\it for a fixed cell geometry}, are:
\begin{equation}
\Pi_{{\bm q}j}(\omega)=
\Pi_{{\bm q}j}^{(T)}(\omega) + \Pi_{{\bm q}j}^{(L)}(\omega) + 
\Pi_{{\bm q}j}^{(B)}(\omega),
\label{eqn1}
\end{equation}
which correspond, respectively, to the tadpole (T), loop (L), and bubble (B)
diagrams of Fig.~\ref{fig1}, whose calculation is described in
{\it e.g.} in Refs.~\cite{manybody,Menendez}.

Let us consider the interatomic potential energy
${\cal E}^{tot}(\{u_{s\alpha}({\bm R}_i)\})$,
where $u_{s\alpha}({\bm R}_i)$ is the
displacement from the equilibrium position of the $s$-th atom in the crystal
cell identified by the lattice vector ${\bm R}_i$ along the $\alpha$
Cartesian coordinate. We define
\begin{eqnarray*}
V^{(n)}({\bm q}_1j_1,...,{\bm q_n}j_n)=\frac{\partial^n {\cal E}^{cell}}
{\partial u_{{\bm q}_1j_1}...\partial u_{{\bm q}_nj_n}},
\end{eqnarray*}
where ${\cal E}^{cell}$ is the energy per unit cell and the adimensional
quantity
$u_{{\bm q}j}$ is defined by
$u_{{\bm q}j}=\frac{1}{N}\sum_{i,s,\alpha} \sqrt{\frac{2M_{s}\omega_{qj}}
{\hbar}}\,
v_{s\alpha}({\bm q}j) u_{s\alpha}({\bm R}_i) e^{-i{\bm q}\cdot{\bm R}_i}$,
$v_{s\alpha}({\bm q}j)$ being the orthogonal phonon eigenmodes normalized
on the unit cell, $M_s$ the atomic mass, and $N$ the number of ${\bm q}$-points
describing the system (or unit cells). Using this definition $V^{(n)}$ is a
intensive quantity, independent of $N$, at any order $n$.
The evaluation of the three diagrams for the $j$ zone center mode
(${\bm q}={\bm 0}$) yields:
\begin{widetext}
\begin{eqnarray}
\nonumber
\Pi_{{\bm 0}j}^{(T)}(\omega)&=&
-\frac{1}{N\hbar^2}\sum_{{\bm q},j_1,j_2}
\frac{V^{(3)}({\bm 0}j,{\bm 0}j,{\bm 0}j_1)V^{(3)}({\bm 0}j_1,-{\bm q}j_2,{\bm q}j_2)}{\omega_{{\bm 0}j_1}}
(2n_{{\bm q}j_2} + 1) \\
\nonumber
\Pi_{{\bm 0}j}^{(L)}(\omega)&=&
\frac{1}{2N\hbar}\sum_{{\bm q},j_1}
V^{(4)}({\bm 0}j,{\bm 0}j,-{\bm q}j_1,{\bm q}j_1)
(2n_{{\bm q}j_1}+1) \\
\nonumber
\Pi_{{\bm 0}j}^{(B)}(\omega)&=&
-\frac{1}{2N\hbar^2}\sum_{{\bm q},j_1,j_2}
|V^{(3)}({\bm 0}j,-{\bm q}j_1,{\bm q}j_2)|^2
\left(
\frac{2(\omega_{{\bm q}j_1}+\omega_{{\bm q}j_2})(1+n_{{\bm q}j_1}+n_{{\bm q}j_2})}
     {(\omega_{{\bm q}j_1}+\omega_{{\bm q}j_2})^{2}-(\omega+i\delta)^2} +
\frac{2(\omega_{{\bm q}j_1}-\omega_{{\bm q}j_2})(n_{{\bm q}j_2}-n_{{\bm q}j_1})}
     {(\omega_{{\bm q}j_2}-\omega_{{\bm q}j_2})^{2}-(\omega+i\delta)^2}
\right),\\
\label{eqn2}
\end{eqnarray}
\end{widetext}
where $\sum_{\bm q}$ is a sum on the Brillouin zone (BZ),
$n_{{\bm q}j}$ is the Bose-statistics occupation of the
phonon mode $({\bm q}j)$, and $\delta$ is a small positive number.
Interpreting the three diagrams of Fig.~\ref{fig1} in terms of
scattering between phonons, the self-energy $\Pi_{{\bm q}j}(\omega)$ can thus
be expressed as a sum, over the BZ, of the phonon
scattering coefficients $V^{(n)}$.

Eqs.~\ref{eqn1},\ref{eqn2} are valid at fixed cell geometry.
In general, the dependence of $\Pi_{{\bm q}j}(\omega)$ on the
temperature depends on further terms related to the thermal expansion of the
lattice.
However, if one can determine the lattice thermal expansion independently
({\it e.g.} experimentally), the shift at a given temperature can be obtained
computing the harmonic frequency at the lattice parameters corresponding to
that temperature.
While only the $\Pi^{(B)}$ term in Eq.~\ref{eqn1} has an imaginary
component ({\it i.e.} is providing a contribution to the line broadening,
which we already calculated and discussed in Ref.~\cite{Shukla}),
all three diagrams have real part and contribute to the shift.
Due symmetry, the $\Pi^{(T)}$ term is zero~\cite{tadpole} in MgB$_2$, and the
frequency shift can, thus, be decomposed in
three- and four-phonon scattering contributions
$\Delta_{{\bm 0},j}^{(3)}=\mathcal{R}_e [\Pi_{{\bm 0}j}^{(B)}
(\omega_{{\bm 0}j})]$ and
$\Delta_{{\bm 0},j}^{(4)}=\mathcal{R}_e [\Pi_{{\bm 0}j}^{(L)}
(\omega_{{\bm 0}j})]$.

Several authors used the frozen-phonon approach to determine MgB$_2$
frequency shift~\cite{Choi,Kortus,Liu,Yildirim,Kunc}. This approach is based on
the study of the energy profile ${\cal E}_{{\bm q=0},j}(\lambda)$ obtained
displacing MgB$_2$ atomic positions by a value proportional to $\lambda$, along
the eigenvector of mode $j$ at ${\bm q=0}$.
This approach is valid when: (i) the coupling of the mode $j$ with
different modes is negligible; (ii) the anharmonic interaction does not change
over the BZ, {\it i.e.} when
$V^{(n)}({\bm 0}j,{\bm q}_2j,...,{\bm q}_nj) \sim
V^{(n)}({\bm 0}j,{\bm 0}j,...,{\bm 0}j)~~
\forall{\bm q}_2,...,{\bm q}_n \epsilon {\rm BZ}$.
Only in this non-dispersive regime the behavior of
${\cal E}_{{\bm q=0},j}(\lambda)$ gives quantitative informations
about the shift.

Our electronic structure calculations is performed using a plane-wave basis
set up to a $35$ Ry cutoff, and using the pseudopotential approach
~\cite{pseudop}.
The terms $V^{(3)}$ of Eq.~\ref{eqn2} are computed analytically with the
approach of Ref.~\cite{Lazzeri02} and the $V^{(4)}$ are obtained by
finite differentiation of $V^{(3)}$.
Calculations are repeated both at T=0~K and T=300~K temperature
crystal structures~\cite{nota00}.
First order Hermite-Gaussian smearing of $0.025$ Ry is used.
The dynamical matrices and phonon frequencies are calculated using
the PBE~\cite{PBE} approximation for the exchange and correlation functional
and a well converged $16\times 16\times 16$ Monkhorst-Pack grid
for the electronic BZ integration.
Due to the actual implementation in our code, third and fourth order
interatomic force constants ($V^{(3)}$ and $V^{(4)}$) are obtained using
the LDA~\cite{LDA} functional.
By performing frozen-phonon calculations at $\Gamma$ using both PBE and LDA
we estimate that the total shift obtained by LDA should be slightly larger
($\lesssim 0.5$ meV) than with PBE.

\begin{table}
\begin{ruledtabular}
\begin{tabular}{cccc}
 $\Delta^{(4)}_{ND}$&$\Delta^{(4)}$&$\Delta^{(3)}$&$\Delta^{tot}$ \\
\hline
\multicolumn{4}{c}{T~=~0~K,~~~~~$\omega$~=~$\bm{ 65.14}$~meV}\\
   $ \bm{9.46}$         &  $\bm{6.00}$          & $\bm{ -2.88}$& $\bm{ 3.12}$          \\
         $\bm{(+15\%)}$ &       $\bm{(+9.2\%)}$  & $\bm{(-4.4\%)}$&       $\bm{(+4.8\%)}$ \\
&&&\\
         8.91           &       7.18            &       -2.99  &       4.19            \\
              (+14\%)   &            (+11\%)    &      (-4.6\%)  &            (+6.4\%)   \\
\hline
\multicolumn{4}{c}{T~=~300~K~~~~~$\omega$~=~$\bm{ 64.51}$~meV}\\
 $\bm{ 10.68}$          & $\bm{ 6.52}$          & $\bm{ -3.03}$  & $\bm{ 3.50}$          \\
         $\bm{(+16\%)}$ &       $\bm{(+10\%)}$  & $\bm{(-4.7\%)}$&       $\bm{(+5.4\%)}$ \\
&&&\\
       10.01            &       7.77            &       -3.12    &       4.65            \\
              (+15\%)   &            (+12\%)    &      (-4.8\%)  &            (+7.2\%)   \\
\end{tabular}
\end{ruledtabular}
\caption{
Phonon frequency shift (meV) of the E$_{2g}$ mode at the BZ center
($\Gamma$ point). $\Delta^{(3)}$ ($\Delta^{(4)}$) is the shift due to the
three(four)-phonon scattering contribution. $\Delta^{tot}$
($\Delta^{tot}=\Delta^{(3)}+\Delta^{(4)}$)
is the total shift. $\Delta^{(4)}_{ND}$ is defined in the text.
$\omega$ is the unperturbed phonon frequency, and the values in parentheses
are the shift as a percentage with respect to the $\omega$ at the corresponding
temperature (T).
Bold face values are obtained using the $16\times16\times16$
electronic integration grid, the others from a less converged
$14\times14\times8$ grid~\protect\cite{nota01}.}
\label{tab1}
\end{table}

\begin{table}
\begin{ruledtabular}
\begin{tabular}{lcccc}
 &$\omega$&$\Delta^{(4)}$&$\Delta^{(3)}$&$\Delta^{tot}$ \\
\hline
\multicolumn{5}{c}{T~=~0~K}\\
$\Gamma$,E$_{2g}$ & $\bm{ 65.14 }$& 7.18 &  -2.99 &  4.19 (+6.4\%) \\     
A (E$_{2g})$        & $\bm{ 57.50 }$& 7.77 &  -3.10 &  4.67 (+8.1\%) \\
M (E$_{2g})$        & $\bm{ 87.18 }$& 4.12 &  -1.98 &  2.13 (+2.4\%) \\
M (E$_{2g})$        & $\bm{ 93.10 }$& 1.10 &  -1.29 & -0.19 (-0.2\%) \\
&&&&\\
$\Gamma$,E$_{1u}$ & $\bm{ 40.03 }$&  0.48 & -0.24 &  0.25 (+0.6\%) \\
$\Gamma$,A$_{2u}$ & $\bm{ 46.21 }$&  1.45 & -0.36 &  1.09 (+2.4\%) \\
$\Gamma$,B$_{1g}$ & $\bm{ 84.49 }$& -1.02 & -0.16 & -1.18 (-1.4\%) \\
\hline
\multicolumn{5}{c}{T~=~300~K}\\
$\Gamma$,E$_{2g}$ & $\bm{ 64.51 }$& 7.77 &  -3.12 &  4.65 (+7.2\%) \\
A (E$_{2g})$        & $\bm{ 56.88 }$& 8.16 &  -3.50 &  4.66 (+8.2\%) \\
M (E$_{2g})$        & $\bm{ 86.65 }$& 4.65 &  -2.13 &  2.52 (+2.9\%) \\
M (E$_{2g})$        & $\bm{ 92.80 }$& 1.20 &  -1.46 & -0.26 (-0.3\%) \\
&&&&\\
$\Gamma$,E$_{1u}$ & $\bm{ 39.74 }$&  0.76 & -0.30 &  0.46 (+1.2\%) \\
$\Gamma$,A$_{2u}$ & $\bm{ 45.81 }$&  0.20 & -0.64 & -0.45 (-1.0\%) \\
$\Gamma$,B$_{1g}$ & $\bm{ 84.28 }$& -1.03 & -0.21 & -1.24 (-1.5\%) \\
\end{tabular}
\end{ruledtabular}
\caption{Phonon frequency shift (meV) of various modes at the three high
symmetry points $\Gamma$, A, and M of the BZ
(see also the caption of Tab.~\protect\ref{tab1}).
The shifts of this table are obtained using the $14\times14\times8$
electronic integration grid~\protect\cite{nota01}.}
\label{tab2}
\end{table}

In Tab.~\ref{tab1} we show the computed values of $\Delta^{(4)}$ and
$\Delta^{(3)}$ (previously defined) for the E$_{2g}$ mode at $\Gamma$.
To make a comparison with the frozen-phonon approach,
in Tab.~\ref{tab1} we also show $\Delta^{(4)}_{ND}$, namely the shift due to
the four-phonon scattering obtained neglecting the interaction with the other
modes, and assuming a non-dispersive anharmonic potential:
{\it i.e.} imposing in Eq.~\ref{eqn2}
$V^{(4)}(\bm{ 0}j,\bm{ 0}j,-\bm{q}j_1,\bm{q}j_1)=
V^{(4)}(\bm{ 0}j,\bm{ 0}j,\bm{ 0}j,\bm{ 0}j) \delta_{j_1,j}$.
As expected~\cite{Kortus,Liu,Yildirim,Kunc,Boeri},
$\Delta^{(4)}_{ND}$ gives a large positive shift
($+16\%$ of the unperturbed frequency at T=300K), which, however,
is far from being a quantitative estimate of the shift.
In fact, the actual determination of scattering processes all over the BZ, and
the inclusion of the other modes ($j_1\neq j$), reduces
the value of the four-phonon shift ($\Delta^{(4)}$ in Tab.~\ref{tab1}) to
just $+10\%$. Furthermore, the inclusion of the $\Delta^{(3)}$ term
(which has a negative sign) reduces the total shift to just $+5.4\%$.
The main reason of the negative sign of $\Delta^{(3)}$
is that the joint density of phonon states
$\rho(\omega) = \sum_{{\bm q},j_1,j_2} \delta(\omega-\omega_{{\bm q}j_1}
-\omega_{{\bm q}j_2})$ is almost entirely (more than $90\%$)
distributed above the frequency of the $\Gamma$-E$_{2g}$ mode.
This implies that, in the summation done
to obtain ${\cal R}_e(\Pi^{(B)})$ from Eq.~\ref{eqn2} at T$=0$,
the largest part of the terms have a negative value.

To compute the shift at the A and M BZ points (${\bm q}\ne{\bm 0}$) we
use, respectively, a $1\times 1\times 2$, and a $2\times 2\times 1$
super-cell.
Due to the heavier computational effort required, $V^{(3)}$ and $V^{(4)}$
are calculated with less converged electronic integration grids equivalent to
a $14\times 14\times 8$ grid for the $1\times 1\times 1$ cell~\cite{nota01}.
From Tab.~\ref{tab1} one can infer that calculations with this grid provide
a qualitative description of the system.
In Tab.~\ref{tab2} we show the shift, at A and M, of the modes belonging to
the E$_{2g}$ band, and, at $\Gamma$, the shift of modes different than
E$_{2g}$.
It is apparent that the total E$_{2g}$ shifts at $\Gamma$
and A are greater than both the E$_{2g}$ shift at M,
and the shift of the other modes at $\Gamma$.
This is not surprising:
it is related to the strong electron-phonon coupling of the E$_{2g}$ mode
along the $\Gamma$-A direction~\cite{Liu}, and confirms one accepted
interpretation of the MgB$_2$ electronic band structure~\cite{Liu,Kong}.
Moreover, the E$_{2g}$ shift at $\Gamma$ and A are very similar, being 4.65
and 4.66 meV (Tab.~\ref{tab2}), respectively, at room temperature.
This confirms the predicted presence of a uniform shift along
the $\Gamma$-A direction~\cite{Liu,Choi}.

Keeping in mind that our best estimate for the $\Gamma$ shift is just
$+3.5$ meV ({\it i.e.} $+5.4 \%$ of the harmonic value, from Tab.~\ref{tab1})
we expect the anharmonic shift to be of the same order all along $\Gamma$-A.
Such a value is compatible with the room temperature inelastic scattering
X-ray measurements of Ref.~\cite{Shukla}.
The measured E$_{2g}$ dispersion along the $\Gamma$-A direction, from
Ref.~\cite{Shukla}, is in agreement with harmonic model calculations.
Indeed, comparing the experimental E$_{2g}$ frequency at A (59.0 meV)
with our calculation, even including anharmonic effects
($56.88 + 4.66 = 61.5$ meV, using the $14\times 14\times 8$ grid) the
agreement remains well within the experimental error-bar and the typical DFT
error.

Finally, we notice that our calculations predict an almost negligible
dependence of the E$_{2g}$-$\Gamma$ frequency from the temperature (being
$65.14+3.12=68.26$ meV at T$=0$K, and $64.51+3.50=68.01$ meV at T$=300$K).
In fact, the largest part of this shift is due to the coupling with phonons
having a Debye temperature larger than 300K ($\omega > 26$~meV).

In conclusion, we have obtained the anharmonic phonon frequency shift in
MgB$_2$ using density functional theory.
Two ingredients which turned out to be essential for a quantitative description
of the anharmonic E$_{2g}$-mode frequency-shift are:
(i) the explicit calculation of the scattering processes all over the BZ;
(ii) the inclusion in the perturbative expansion of both the three- and
four-phonon scattering contributions, which have opposite sign.
The resulting phonon frequency shift at $\Gamma$ is just +5.4\%
of the harmonic frequency, and is expected to be of the same order along
the $\Gamma$-A direction, in agreement with inelastic X-ray
scattering~\cite{Shukla}.

We are grateful to O.K.Andersen, J.Kortus, I.Mazin, H.J. Choi,
K.M.Rabe, and A.Shukla for useful discussions.
M.C. was supported by a Marie Curie Fellowship,
contract No. IHP-HPMF-CT-2001-01185.
Calculation were done at the IDRIS supercomputing centre (Orsay, France),
using the PHONON code: http://www.pwscf.org~.



\begin{thebibliography}{99}
\bibitem{nota00}
We use the room temperature structure of ~\protect\cite{Shukla},
and extrapolating the T=0~K one according to the experimental thermal 
expansion of Ref.~\protect\cite{Jorgensen}.
Third and fourth order derivatives of the energy are computed on a 
$4\times 4\times 2$ mesh of phonon ${\bm q}$-points in the BZ corresponding to
the $1\times 1\times 1$ cell, and Fourier interpolated at the ${\bm q}$-points
of the thicker ($20^3$) grid used for the summation in
Eq.~\protect\ref{eqn2}.
Other computational details are the same as in Ref.~\protect\cite{Shukla}.
\bibitem{nota01}
The $14\times14\times8$ grid is used to compute only $V^{(3)}$ and $V^{(4)}$.
While computing the shifts from Eqs.~\protect\ref{eqn2}, we always use
frequencies calculated using the well converged $16\times16\times16$
electronic integration grid.
\end{thebibliography}
\end{document}